**Reversible Control of In-plane Elastic Stress Tensor in Nanomembranes**

*Javier Martín-Sánchez[1*], Rinaldo Trotta[1], Giovanni Piredda[2], Christian Schimpf[1], Giovanna Trevisi[3], Luca Seravalli[3], Paola Frigeri[3], Sandra Stroj[2], Thomas Lettner[1], Marcus Reindl[1], Johannes S. Wildmann[1], Johannes Edlinger[2] and Armando Rastelli[1*]*

[1] Institute of Semiconductor and Solid State Physics, Johannes Kepler University, Altenbergerstr. 69, A-4040 Linz, Austria

[2] Research Center for Microtechnology, Vorarlberg University of Applied Sciences, 6850 Dornbirn, Austria

[3] IMEM-CNR Institute, Parco Area delle Scienze 37/a, 43124 Parma, Italy

* e-mail: javier.martin_sanchez@jku.at; armando.rastelli@jku.at

The physical properties of materials critically depend on the interatomic distances of the constituent atoms, which in turn can be tuned by introducing elastic strain (stress). While about four decades ago strain was generally regarded as a feature to be avoided in semiconductors,[1] strain engineering is nowadays ubiquitously used, e.g., to enhance the carrier mobility in transistors [2][3] and to achieve lasing action at reduced current densities in heterostructure lasers.[4] For this reason, its potential impact on our society has been compared to that of chemical alloying.[5] Strain can be used not only to enhance specific material/device properties, but also to impart completely new properties to a given material, thus opening the way to previously inaccessible applications. Examples are Ge turning into a direct-bandgap semiconductor suitable for lasers,[6][7][8][9] exciton dynamics tailoring in nanowires induced by strain gradients,[10] graphene electronic states engineering and strain-induced giant pseudo-magnetic fields up to 300 T,[11][12] a topological insulator turning into a semiconductor,[13] linear electro-optical effects in Si,[14] bandgap modulation in atomically thin films [15] or surface chemical and electronic states tuning.[16]

In the above examples strain is used in a static fashion, i.e. its magnitudes and directions are frozen during device fabrication. For some applications and for investigating strain-induced effects in new materials, it would be instead desirable to be able to precisely manipulate the stress state of a material/device during measurement/operation. Examples can be found in a wide range of research areas such as nanophotonics,[17][18][14] electronic and energy conversion applications,[19] photovoltaics,[20] spintronics,[21] topological insulators [13] or graphene [11] and other 2-D materials.[22] Different approaches are being pursued to exert non-hydrostatic stresses on materials, such as three-points bending,[23] piezoelectric stacks actuators,[24] 2-D films deposition on pre-patterned substrates,[25] growth or integration of functional films on piezoelectric substrates [26] and surface acoustic waves. [27]

For anisotropic materials, the general case for crystalline substances, it would be desirable not only to control the strain magnitude, but also its direction and anisotropy. Here, we address this issue for the first time by introducing a novel class of strain-actuators capable of controlling the three components of the in-plane stress tensor in thin films. The functionality of the actuators is demonstrated by engineering the strain state of a semiconductor nanomembrane made of GaAs. We show that arbitrary stress configurations (isotropic and anisotropic biaxial as well as uniaxial with different directions) can be produced on the very same area of the sample by properly changing the values of three voltages applied to the actuator.

We present now the working principle and experimental implementation of our actuator. For in-plane stress, the stress state at a given position is completely determined by three components of the stress tensor ($S_{xx}$, $S_{yy}$, $S_{xy}$) or, equivalently, by the principal major ($S_1$) and minor ($S_2$) stresses and the angle ($\theta_{S1}$) between the major stress axis and a reference direction.

For a given stress configuration, also the strain state is fully determined, as it is related to the stress by the compliance tensor of the material. Full control of the stress state means that the above three parameters can be tuned independently. To achieve this control in practice, the actuator should feature *three independent "tuning knobs"*, capable of generating three in-plane stress configurations with directions that are not all parallel to each other. In order to obtain compact actuators, which can operate in different environments (e.g. vacuum, low temperatures, magnetic fields), we build up our devices starting from 300-μm-thick monolithic $[Pb(Mg_{1/3}Nb_{2/3})O_3]_{0.72}[PbTiO_3]_{0.28}$ (PMN-PT) piezoelectric substrates, which can conveniently transduce electric voltages into mechanical deformations. As we need three independent "tuning knobs", we micro-machine the actuators by femtosecond laser cutting into three legs, whose deformation is controlled via three independent voltages.

The device consists of a 400-nm-thick GaAs nanomembrane bonded on the micro-machined actuator by gold thermo-compression bonding. GaAs was chosen to characterize the actuators and demonstrate the capabilities, as it is a well-known direct-bandgap semiconductor with bright photoluminescence (PL) and known response to deformations.[28, 29, 30] For further details about sample growth and device fabrication, see Supporting Information. **Figure 1**a shows a sketch and a microscope picture of an actuator featuring three legs with about 300×300 μm$^2$ cross-section, 1500 μm length, and radially arranged around the region where the GaAs nanomembrane to be stressed is located. Three voltages ($V_1$, $V_2$, $V_3$) are applied between the bottom of the legs and the top, which is set to ground. Such voltages produce vertical electric fields across the piezoelectric legs, leading to their in-plane deformations, which are transferred to the bonded nanomembrane. The sign of the stress (tensile/compressive) exerted on the membrane is controlled by the voltage polarity (+/-) while the stress magnitude is tuned by the voltage magnitude. Membrane buckling limits the range of compressive stresses which can be applied with the actuator. Thus, most of the measurements were performed in the tensile regime.

We discuss now the method used to retrieve the stress state of a nanomembrane and its relation to the voltages applied to the actuator. The stress state in the gap between the micro-machined legs (white spot in Figure 1a) is fully retrieved using polarization-resolved micro-PL spectra. In this scheme, the GaAs PL emission is used as a local strain gauge. The PL signal is collected along the direction perpendicular to the nanomembrane surface as a function of linear-polarization angle. In order to retrieve the stress state in the nanomembrane, we solve first the corresponding direct problem, consisting of calculating the expected spectra for a given stress configuration, by 8-band k.p theory with stress introduced via the Pikus-Bir Hamiltonian and optical properties treated with the dipole approximation. The Hamiltonian, elastic constants and deformation potentials are taken from Ref. [29] (see also Table S1 in Supporting Information). Since the minimum of the conduction band (CB) and the maximum of the valence band (VB) of GaAs are both located at the Γ point of the Brillouin zone, we diagonalize the Hamiltonian at such point to obtain the single-particle energies of electrons and holes as well as the corresponding eigenstates. The latter are used to extract the transition probabilities as a function of polarization direction. In absence of strain, light-hole (LH) and heavy-hole (HH) bands are degenerate at the Γ point, so that a single peak is observed in the PL emission. An in-plane stress (non-hydrostatic) breaks the crystal symmetry and removes the VB degeneracy, leading to two emission peaks (see, e.g., Figure 1b). As measurements were performed at 10 K, these peaks are ascribed to the radiative recombination of free excitons consisting of electrons in the spin-degenerate CB and holes in the two strain-split VBs. In the model, excitonic effects are included by a simple renormalization of the bandgap energy, i.e. by taking the experimentally measured emission energy of unstrained GaAs (1.5148±0.0001 eV). In general, i.e. for non-purely biaxial stress, the two VBs consist of an admixture of HH and LH bands. We will thus refer hereafter to the low- and high-energy PL components as $E_1$ and $E_2$, respectively. Due to band mixing, the light as collected along the perpendicular direction to the nanomembrane – which is unpolarized for unstrained or biaxially strained material – becomes elliptically polarized, as shown in the polar plots of Figure 1b. In turn, the polarization stems from the anisotropy of the Bloch wave-functions of the corresponding mixed hole states, as illustrated by the probability density distributions shown on the right panel of Figure 1b. Using the theoretical framework described above, we find that the low/high energy component is polarized along a direction close to the minor/major stress axes and the stress state of the nanomembrane, fully characterized by $S_1$, $S_2$, and $\theta_{S1}$ (measured here with respect to the [110] crystal direction of GaAs), is univocally encoded in the energies $E_1$ and $E_2$ and the polarization angle ($\phi$) of the $E_1$ peak. Since the problem cannot be inverted analytically, we use these observables as input data for a non-linear least-square minimization algorithm, which retrieves the $S_1$, $S_2$, and $\theta_{S1}$ combination which minimizes the deviations between measured and calculated $E_1$, $E_2$ and $\phi$. The first guess for the fit is obtained from a semi-analytical solution of the Pikus-Bir Hamiltonian without inclusion of the split-off band. Self-consistency and uniqueness of the solution were thoroughly checked.

We now focus on the relation between applied voltages and stress state of the nanomembrane. In the linear regime, where the stress exerted by the actuator is proportional to the applied voltages, the total stress state can be written as:

$$\bar{S}(\bar{V}) = \bar{S}_{pre-stress} + \bar{s}_{induced}(\bar{V}) = \bar{S}_{pre-stress} + \bar{\bar{T}} \cdot \bar{V} \qquad (1),$$

where $\bar{S} = (S_{xx,}S_{yy,}S_{xy})$ are the relevant components of the stress tensor in Voigt notation, $\bar{S}_{pre-stress}$ the corresponding values at zero applied voltages, $\bar{s}_{induced}$ the stress induced by the actuator, and $\bar{\bar{T}}$ is a 3×3 "transfer matrix", with each row containing the proportionality coefficients between the voltages $\bar{V} = (V_1, V_2, V_3)$ applied to the three legs and the corresponding component of the stress tensor (see Supporting Information).

The pre-stress, induced during membrane bonding, piezo poling, and sample cooling, leads to a splitting of the emission already at zero applied voltage, as shown in the example of **Figure 2**a. The corresponding pre-stress is in this case anisotropic biaxial compressive with principal stresses $S_1$=-8±1 MPa, $S_2$=-78±2 MPa, and direction $\theta_{S1}$ =110±1°. If we now ramp the voltage of one leg ($V_3$ in Figure 2a), the two peaks show an anticrossing pattern, with a minimum splitting of 2.5 meV, which corresponds to an induced stress $\bar{s}_{induced}$ having principal components $s_1$=40 MPa, $s_2$=-10 MPa and direction $\theta_{s1}$ =1°. The anticrossing is due to interaction between the HH and LH bands and indicates that the pre-stress cannot be canceled with one leg alone. In order to recover the crystal symmetry (unstrained condition), the actuator should in fact induce a stress $\bar{s}_{induced}$, which exactly compensates $\bar{S}_{pre-stress}$. By proper tuning of $V_1$and $V_2$, the level degeneracy (unstrained GaAs) can be fully recovered as shown in Figure 2b and by polarization-resolved measurements at the crossing point. Figure 2b shows the evolution of the $E_1$-$E_2$ splitting vs $V_3$ as well as the polarization-resolved PL polar plots for $E_1$ component as the stress is swept through the crossing point. A clear 90° rotation of the polarization angle (i.e. major/minors stress axis) is observed. We point out that the same result was obtained at different points of the membrane, which are characterized by slightly different pre-stress states, already showing that arbitrary stress configurations can be produced by the actuator.

We now move a step forward and show how a desired stress state can be "programmed" at a given position of the nanomembrane. To this aim, we first calibrate the actuator, i.e. retrieve pre-stress and transfer matrix, by best fitting the stress values obtained for a relatively small set of voltages with **Equation 1** (see Figure S2 in Supporting Information). The stress induced by the actuator is then calculated as $\bar{s}_{induced} = \bar{\bar{T}} \cdot \bar{V}$. The actuating direction of each leg is given by the direction $\psi_{s1}$ (fixed) of the principal stress (tensile) obtained for any positive voltage applied to the leg. For the example discussed here, the angles are 13±2°, 100±2° and 178±2° for leg1, leg2 and leg3, respectively (see also Figure S4 in Supporting Information). We attribute the deviation with respect to the angles expected from the design (30º, 90º and 120º) to inhomogeneities in the bonding of the nanomembrane on each of the legs and to the measurements having been performed at a slightly different position from the center. It is important to note that this deviation does not affect the working principle of the actuator, which only requires the three induced in-plane stresses to be non-parallel.

After calibration, we designed several experiments to unequivocally demonstrate the capabilities of the device to exert in-plane stresses on demand. In a first experiment we calculated, by inverting Equation 1, the voltages required to achieve a tensile *uniaxial stress at a fixed angle* ($\psi_{s1}$=15° with respect to the [110] crystal direction of GaAs) and a magnitude linearly varying from 0 to 100 MPa (see Figure S3 in Supporting Information). The voltage ramps were repeated three times to collect PL at three different values of the polarization angle of 0°, 45° and 90°. The different intensities observed on both PL components are due to the rotation of the stress field (i.e. light polarization) as the voltages are varied. The results are shown in the left panels of **Figure 3**. Both the energy shifts and the evolution of the peak intensities are in very good agreement with the calculated spectra (see right panels of Figure 3), indicating fine and predictable stress magnitude tuning over the full range. (In order to facilitate the comparison, the full width at half maximum (FWHM) value for the calculated peaks is set to ~1 meV, similar to the experimental data). Only a small deviation is observed for the largest stress value (100 MPa), which we attribute to non-linear response of the piezoelectric substrate at high electric fields (not included in our model). We note that the agreement between experimental and simulated data persists in all three measurement sets, indicating excellent reproducibility of the stress configuration.

Second, we demonstrate control over the induced in-plane stress direction. The panels in **Figure 4**a show the comparison between target and experimentally obtained ($s_1$, $s_2$, $\psi_{s1}$) values for different induced stress field configurations. In particular, we applied deliberate uniaxial stresses with $s_1$=20 and 40 MPa along different directions $\psi_{s1}$=0° and 45°, isotropic biaxial stresses ($s_1$=$s_2$=20 and 40 MPa) and anisotropic biaxial stresses ($s_1$=-20 MPa, $s_2$=0 MPa and $\psi_{s1}$=0° and 90°). In all cases, the target and experimental data are in excellent agreement within the experimental error which clearly confirmed the validity of the proposed device to exert induced stress fields on demand.

The tuning capabilities of every actuator can be conveniently summarized by plotting the maximum tuning range (achievable within a "safe" range of electric fields across the piezoelectric material) of the induced hydrostatic stress ($s_1$+$s_2$) against stress anisotropy ($s_1$-$s_2$) and angle of the principal major stress $\psi_{s1}$. Figure 4b shows, in cylindrical coordinates, such a tuning range for the device shown in Figure 1a and electric fields ranging from -2 to 33 kV/cm. The minimum and maximum hydrostatic stress values achievable for all possible stress anisotropies and directions are depicted. Remarkably, the stress anisotropy can be tuned up to about 50 MPa for any direction and the hydrostatic stress values can be tuned by up to 150 MPa. The negative voltages (i.e. compressive induced stress) have been limited to modest values in order to avoid possible non-linear effects due to curling of the nanomembrane. The different angular response of the device can be well explained by considering the actuating directions of the legs.

In conclusion, we have demonstrated a new class of compact strain-actuators, which allow the three components of the in-plane stress tensor in a nanomembrane to be independently and reversibly controlled. Their functionality is demonstrated by "programming" arbitrary stress states in a semiconductor layer, whose light emission is used as a local and sensitive strain gauge. While we have focused on GaAs membranes at low temperature and vacuum environment, the actuator can be operated also at room temperature and environment conditions as well as in magnetic fields, making it as an ideal platform to investigate strain-induced effects in new materials such as metal dichalcogenides and topological insulators. Preliminary experiments on different actuator designs (see Supporting Information) show that higher stress values can be already obtained with the same tunability demonstrated here. It is worth mentioning that Finite Element Simulations show enhanced stress magnitudes up to 1 GPa at room temperature working conditions.[18]

## Experimental section

*Photoluminescence spectroscopy:* We use a standard confocal micro-PL setup equipped with a X-Y stage allowing spatial positioning of the laser spot with a resolution well below 1μm. A HeNe continuous-wave laser is used for excitation ($\lambda$=633 nm). The laser beam is focused on the GaAs membrane by using a 10x objective with a numerical aperture value of 0.28. The experiments are performed at 10K using a He flow cryostat. PL is analyzed with a spectrometer coupled to a Si CCD using a 600 grooves/mm grating. In order to perform the linear-polarization-resolved measurements, we use a half-lambda wave plate combined with a fixed linear polarizer with its transmission axis aligned along the [110] GaAs crystallographic direction with a resolution of ±5 degrees.

*Femtosecond laser cutting:* A commercial 3D-Micromac laser micromachining system is used. The femtosecond laser is operated with a 350 fs pulse duration, 25 KHz repetition rate and 6 μJ pulse energy. For a better profile quality, the laser is focused down to a spot size of 5 μm.

## Supporting Information

Supporting Information is available from the Wiley Online Library or from the author.

**Acknowledgements**

We thank A. Hallilovic for valuable technical support. E. Lausecker and G. Katsaros are acknowledged for fruitful discussions. The work was supported financially by the European Union Seventh Framework Program 209 (FP7/2007-2013) under Grant Agreement No. 601126 210 (HANAS) and the AWS Austria Wirtschaftsservice, PRIZE Programme, under Grant No. P1308457.

Received: ((will be filled in by the editorial staff))
Revised: ((will be filled in by the editorial staff))
Published online: ((will be filled in by the editorial staff))

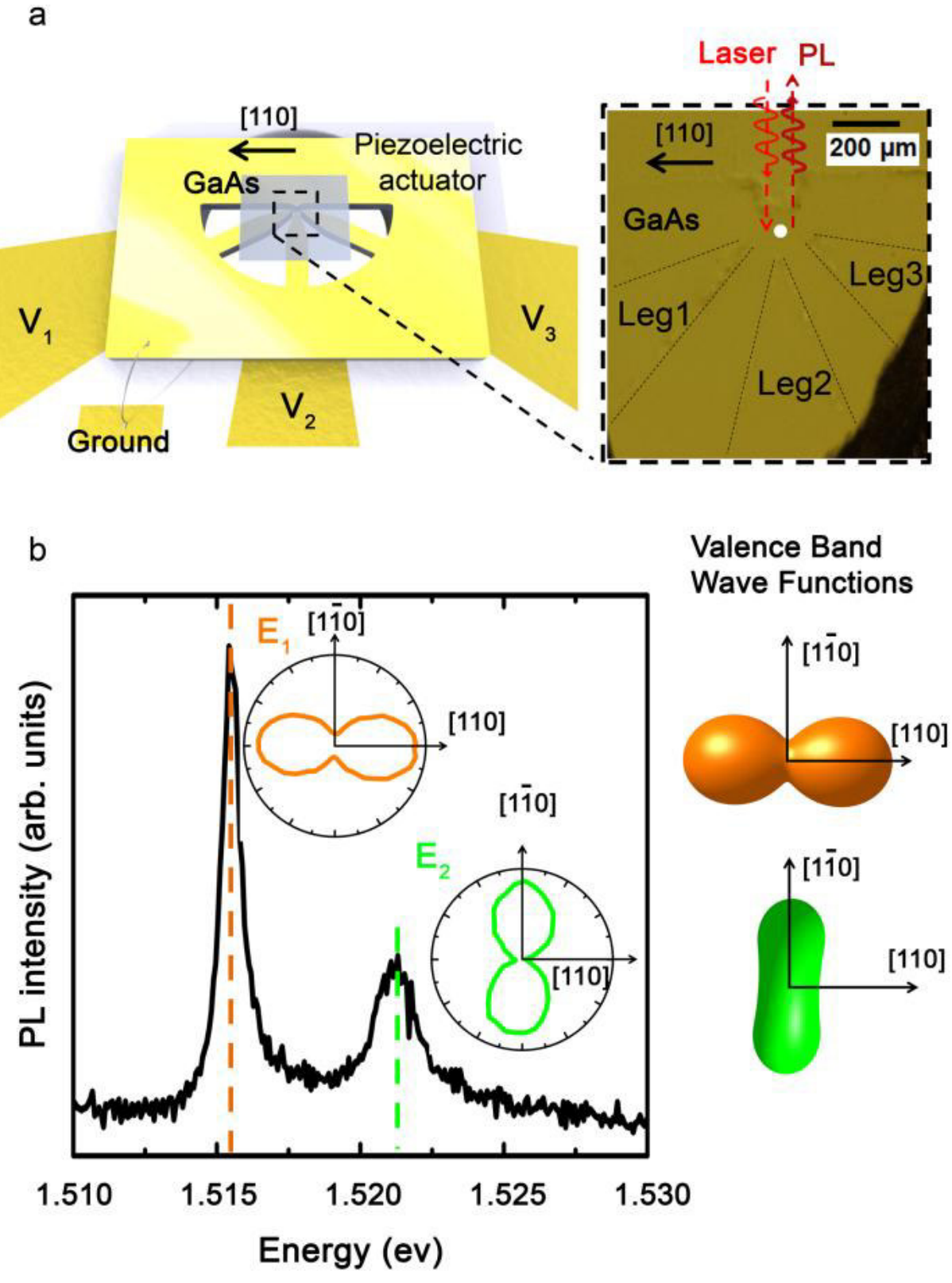

**Figure 1.** Micromachined piezoelectric actuator for full control of the in-plane stress. (a) schematic representation (left) and optical microscope picture (right) of the 3-legged device with a GaAs nanomembrane bonded on it. Three independent voltages ($V_1$,$V_2$,$V_3$) applied

across the legs, induce tunable deformations in the nanomembrane. The white circle on the picture shows the area where linear-polarization-resolved PL measurements were performed. (b) Typical PL spectrum at 10K of GaAs nanomembrane under strain. The anisotropic strain lifts the degeneracy between the valence bands and leads to two emission peaks at energies $E_1$ and $E_2$. The intensity of the two peaks as a function of polarization angle is shown on the right. The energies $E_1$, $E_2$ and polarization of the $E_1$ component are used to retrieve the stress in the nanomembrane through our theoretical model. The calculated probability density functions for LH and HH valence band states are shown on the right.

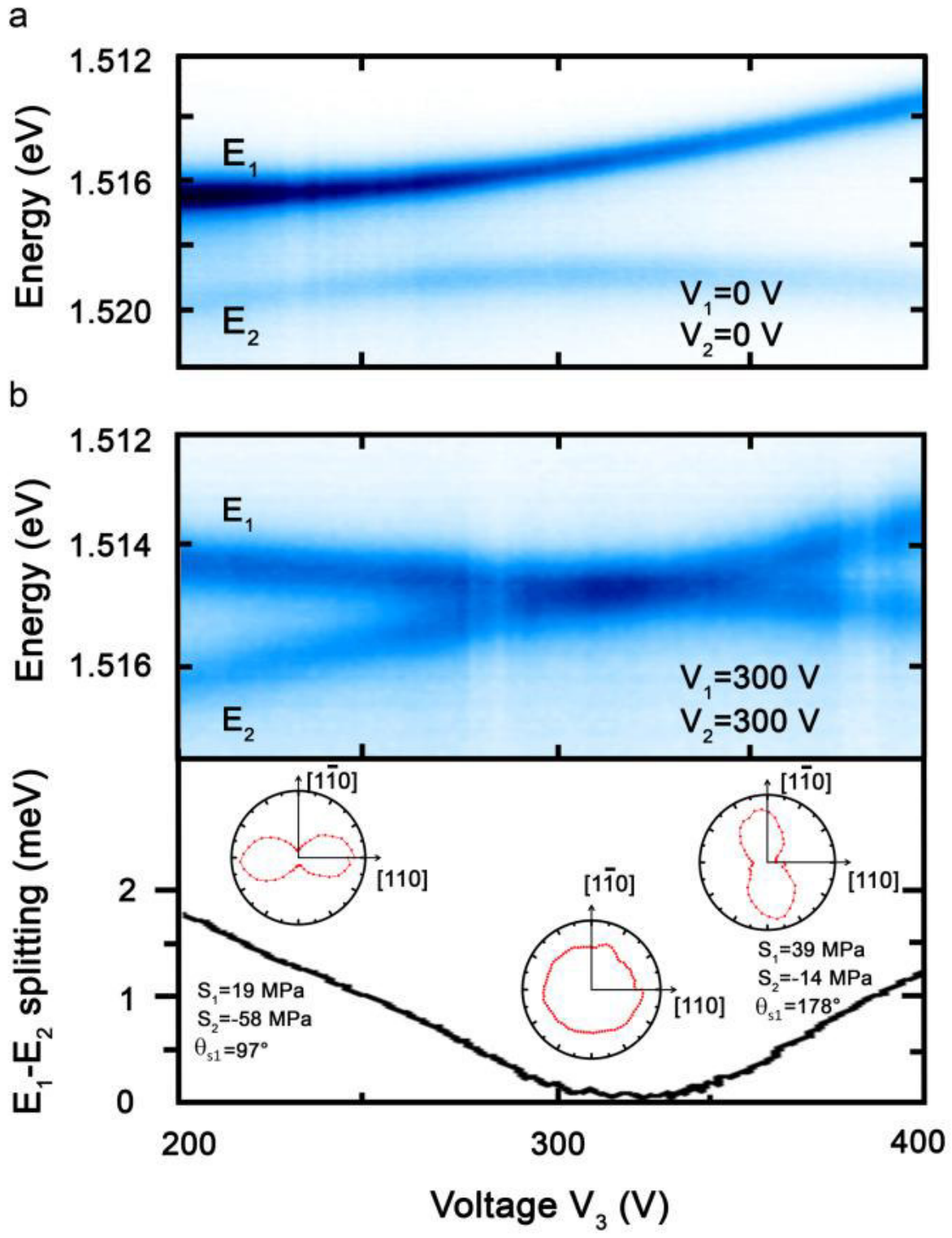

**Figure 2.** Tailoring of the total stress field in a nanomembrane. (a) Color-coded PL spectra of GaAs as a function of the voltage $V_3$ applied to one leg of the actuator. The splitting at zero voltage is due to pre-stress induced during the fabrication process. As the anisotropic biaxial stress induced by the tuning leg has principal directions non-parallel to the pre-stress, the two emission peaks anti-cross as the voltage is tuned. (b) The same as (a) but for a proper choice

of $V_1$ and $V_2$ which aligns the stress principal directions along the tuning direction of leg3. In this case $V_3$ can be used to recover the crystal symmetry (i.e. unstrained GaAs), indicated by the point where the two emission peaks cross. The polarization angle for $E_1$ PL component (i.e. total major/minor stress axis) rotates by 90° when the stress is tuned through the crossing point where LH-HH degeneracy is recovered.

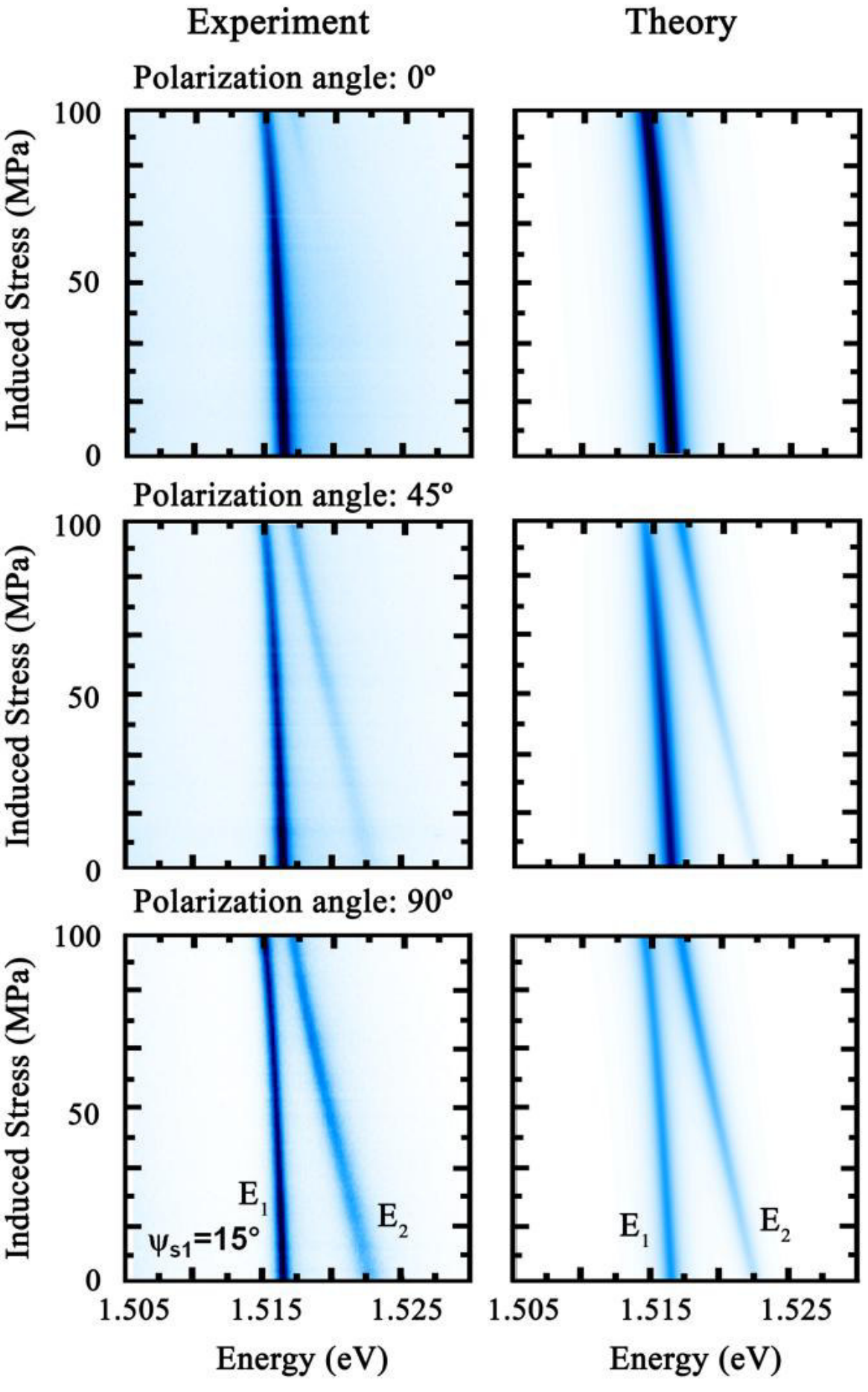

**Figure 3.** Uniaxial stress with fixed direction and variable magnitude. Color-coded PL spectra collected at three different polarization directions (0°, 45° and 90°) while sweeping the actuator voltages in order to induce a purely uniaxial stress of up to 100 MPa along a fixed

direction $\psi_{s1}$ =15° with respect to GaAs [110] crystallographic direction. Experimental data (left) and simulations (right) show excellent agreement demonstrating fine control of the induced stress magnitude and direction.

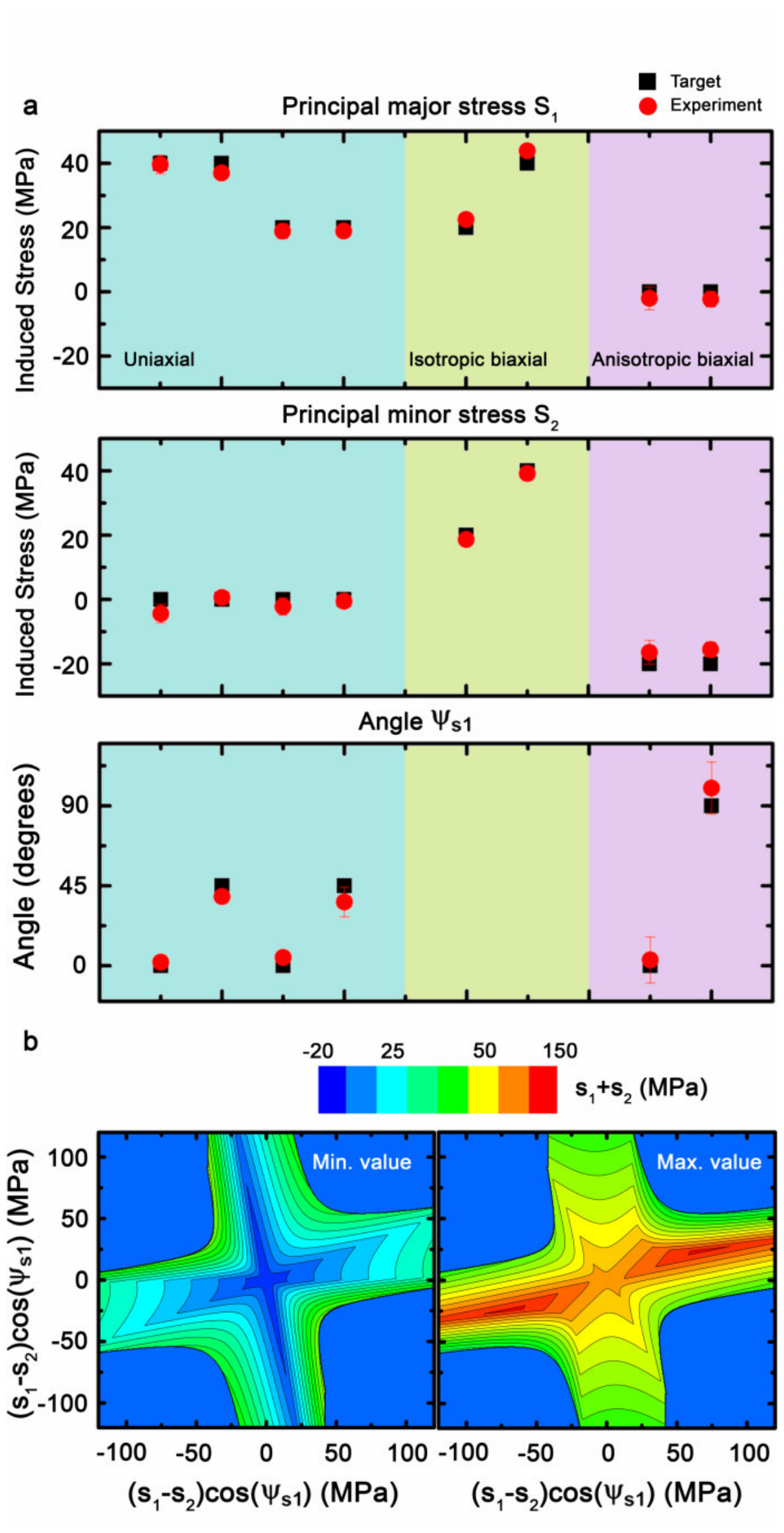

**Figure 4.** Full tuning of the induced in-plane stress field anisotropy. (a) Predicted (square) and experimentally (circles) obtained induced stress $s_1$ (upper panel), $s_2$ (middle panel) and angle of the principal major stress $\psi_{s1}$ (lower panel) for uniaxial, isotropic biaxial and anisotropic biaxial stress field configurations. (b) Tuning range of the device to exert deliberate induced stress fields for electric fields ranging from -2 to 33 kV/cm. The induced hydrostatic stress ($s_1$+$s_2$) against stress anisotropy ($s_1$-$s_2$) and angle $\psi_{s1}$ is plotted in cylindrical coordinates for achievable maximum and minimum hydrostatic stress values.

# Supporting Information

## Reversible Control of In-plane Elastic Stress Tensor in Nanomembranes

*Javier Martín-Sánchez[1*], Rinaldo Trotta[1], Giovanni Piredda[2], Christian Schimpf[1], Giovanna Trevisi[3], Luca Seravalli[3], Paola Frigeri[3], Sandra Stroj[2], Thomas Lettner[1], Marcus Reindl[1], Johannes S. Wildmann[1], Johannes Edlinger[2] and Armando Rastelli[1*]*

* e-mail: javier.martin_sanchez@jku.at; armando.rastelli@jku.at

## Sample processing and device fabrication

The fabrication of the 3-legged micro-machined $[Pb(Mg_{1/3}Nb_{2/3})O_3]_{0.72}[PbTiO_3]_{0.28}$ PMN-PT piezoelectric actuator is performed using a femtosecond laser with 350 fs pulse duration, 25 KHz repetition rate and 6 µJ pulse energy. The laser is focused down to a spot size of 5 µm. Afterwards, the actuator is coated with a [Cr(10 nm)/Au(100 nm)] thin film deposited by electron-beam and thermal evaporation, respectively, on side A of the actuator (**Figure S1**a). The same metallic coating is deposited on side B where three contacts are defined by conventional optical lithography and lift-off processes (Figure S1b). The sample containing the GaAs nanomembrane is bonded on the micro-machined actuator by gold thermo-compression bonding (temperature=300°C; pressure=1 MPa; time=30 min) using flip-chip bonding technique (Figure S1c). In order to release the nanomembrane on the actuator, the 400-µm-thick GaAs substrate is back-etched by selective wet chemical etching in a three-step process: i) 1 hour non-selective etching in $H_3PO_4$:$H_2O_2$ to remove most of the substrate; ~50-µm-thick GaAs layer is left after this step; ii) selective etching in citric acid:$H_2O_2$ for final removal of the substrate up to the $Al_{0.7}Ga_{0.3}As$ etch-stop layer (Figure S1d); iii) HF removal of $Al_{0.7}Ga_{0.3}As$ layer (Figure S1e). The actuator is then contacted (side B) with conductive silver paint on a home-made AlN ceramic chip-carrier where gold paths are previously patterned for the application of the voltages ($V_1$,$V_2$,$V_3$) on each of the legs (see Figure 1a in the manuscript). The other side (side A) is wired with wedge bonding technique and set to ground. The AlN chip-carrier features a hole at its center in order to allow free displacements of the legs and provides good thermal contact with the underlying cold finger for measurements at cryogenic temperatures.

**Data analysis**

After collecting polarization-resolved PL data (with typical angular resolution of 4°) for various sets of voltages we fit the spectra with a double Lorentzian function to extract the emission energies and polarization direction ($E_1,E_2,\phi$). These parameters with corresponding uncertainties are then used to extract the corresponding stress configuration using the model described in the text. Uncertainties are propagated using a Monte Carlo approach consisting in repeating the stress estimation for randomly sampled ($E_1,E_2,\phi$) according to the measurement uncertainties. Having obtained a set of ($S_{xx}$,$S_{yy}$,$S_{xy}$) values for a certain set of voltages ($V_1$,$V_2$,$V_3$), we employ Eq. (1) in the main text to obtain the coefficients of the "transfer matrix" ($\bar{\bar{T}}$) and the prestress. **Figure S2** shows the values of ($S_{xx}$,$S_{yy}$,$S_{xy}$) and corresponding linear fit used for the calibration of the actuator discussed in the text. Only slight deviations from the expected linear behavior are observed.

To further confirm the reliability of the results we compared experimental and calculated degree of polarization (DOP). The trends of DOP are well reproduced, but the calculated DOP is systematically higher than the experimental one, which we ascribe to correlation effects not included in the model. This is the reason for choosing the three observables mentioned above. The analysis of the data also allows us to set a stringent lower limit of -2.2 eV to the value of the deformation potential b, consistent with the recommendations of Ref. [30]. In fact some data sets cannot be fitted assuming lower values.

**Calculated voltages for "programmed" stress sweep**

**Figure S3** shows the voltages calculated through the presented linear model to be applied on each of the legs in order to exert dynamically a tensile stress for a fixed direction ($\psi_{s1}$=15°) (Figure 3 in the main manuscript). The magnitude of s1 is swept from 0 to 100 MPa.

**Validation of the calculated legs actuating directions**

The direction of the principal major stress is directly related to the polarization angle of the high energy PL component ($E_2$) and therefore, it can be obtained from the experimental polarization-resolved PL polar plots. Once the pre-stress is cancelled (i.e. LH-HH degeneracy is recovered), pulling with individual legs tends to align the polarization angle of $E_2$ along the actuating directions as shown in **Figure S4**. This further confirms the validity of the linear model based on the "transfer matrix".

**6-legged piezoelectric actuator**

In this work, we have discussed a particular 3-legged device capable of tuning the stress state of a nanomembrane. This specific design has been chosen because of its simplicity and relatively easy processing. However, in case a higher stress magnitude is needed with the same full tunability (i.e. magnitude, direction and anisotropy control of the stress field), we propose another piezoelectric actuator that features 6-legs, as shown schematically in **Figure S5**a. The working principle of the device is similar to the 3-legged device. In this case, the set of voltages ($V_1$,$V_2$,$V_3$) are applied on pair of legs one in front of each other. In this case slight lateral displacements of the nanomembrane on the area of measurement can be fully avoided. The PL measurements are performed on the suspended nanomembrane at the center of the gap between the legs. Similarly to the 3-legged device, the transfer matrix can be calculated and therefore, given stress states in the nanomembrane can be predicted. Figure S5b shows the GaAs PL emission shift while applying dynamically "programmed" voltages in order to cancel the pre-stress state in the nanomembrane (i.e. $E_1$-$E_2$ crossing is obtained). Figure S5c shows the tuning range of the device plotted in cylindrical coordinates (induced hydrostatic stress ($s_1$+$s_2$) against stress anisotropy ($s_1$-$s_2$) and angle of the principal stress $\psi_{s1}$) for electric fields ranging from -2 to 33 kV/cm. The maximum and minimum achievable hydrostatic stress values are plotted. Interestingly, in this case, the stress anisotropy can be tuned up to about 200 MPa for the full $\psi_{s1}$ angle range with hydrostatic stress values as large as 350 MPa. This means a stress magnitude tunability enhancement of about a factor 2 with respect to the 3-legged device.

It should be noted that PMN-PT piezoelectric material response drops down to about the 25% of the corresponding value at room temperature when working at cryogenic conditions well below 50K (10 K, in the presented work) [1]. In addition, although gold thermo-compression bonding provides a good strain transfer layer between the piezo actuator and the bonded nanomembrane due to the relatively high gold Young modulus (~ 79 GPa), the magnitude of the exerted stress can be reduced with respect to an ideal case due to bonding quality related to the fact that the bonding is done in air and for relatively low pressure values (1 MPa) (i.e. formation of voids in the bonding layer can be obtained).

However, we believe that by improving the bonding quality either by performing gold bonding in vacuum, or using other technological approaches like hard epoxy polymer-based bonding, as well as working at room temperature (usual for most of the applications), the pulling capabilities of the actuator can be significantly improved approaching the theoretically predicted stress field magnitudes [2].

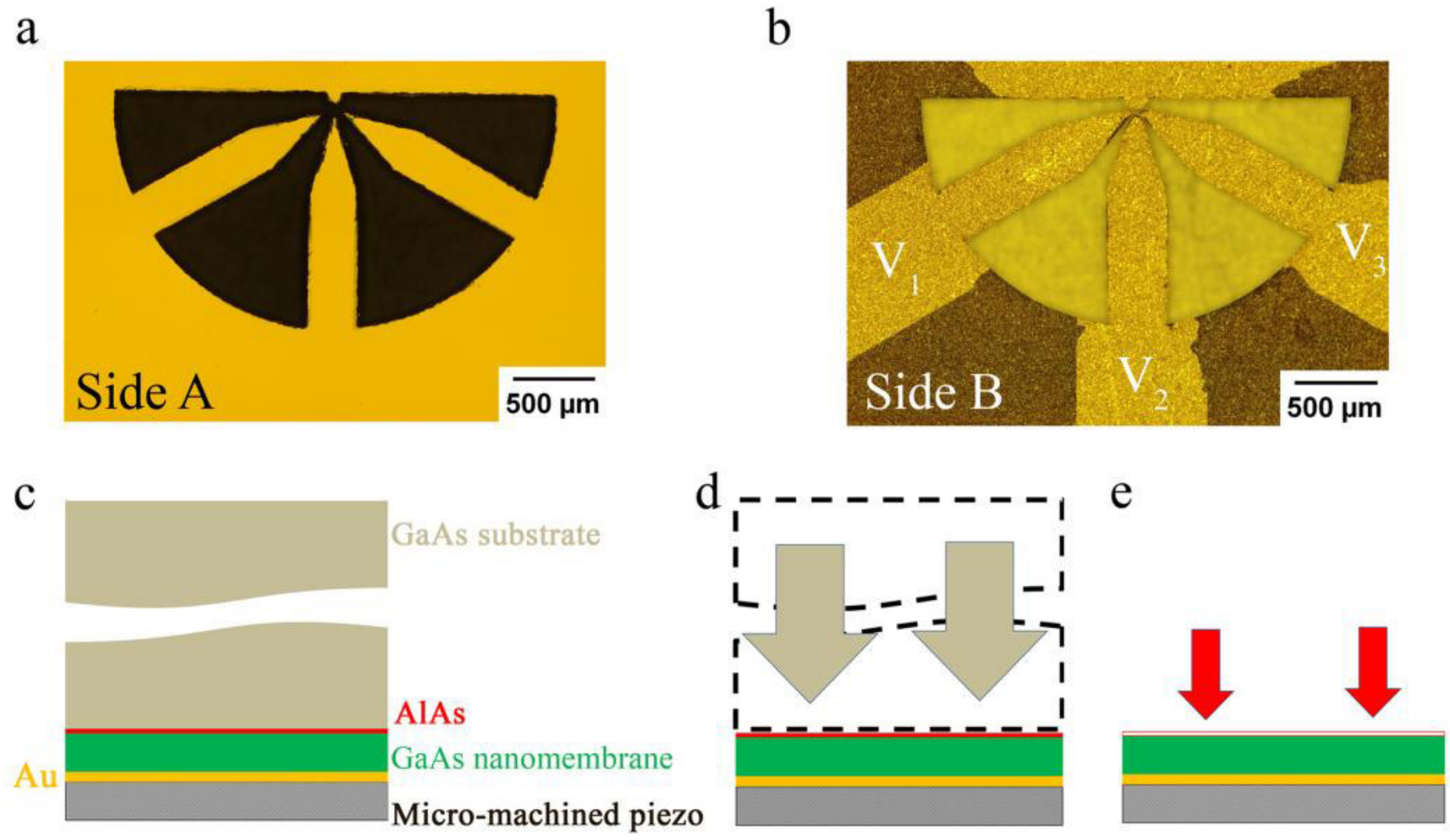

**Figure S1.-** Micro-machined piezoelectric actuator patterning and sample processing. (a) Side A of the actuator is fully Au-coated and set to ground during measurements. (b) Three Au contacts are patterned on side B in order to apply three independent voltages to each of the legs. (c) Schematic representation of the cross-section of the sample [GaAs substrate/ $Al_{0.7}Ga_{0.3}As$ (100 nm)/GaAs nanomembrane (400 nm)] bonded on the micro-machined actuator by Au thermocompression bonding. (d) Schematic representation for the structure obtained after removing the substrate by selective wet chemical etching. (e) Schematic representation of the nanomembrane bonded on the actuator after removal of the $Al_{0.7}Ga_{0.3}As$ sacrificial layer.

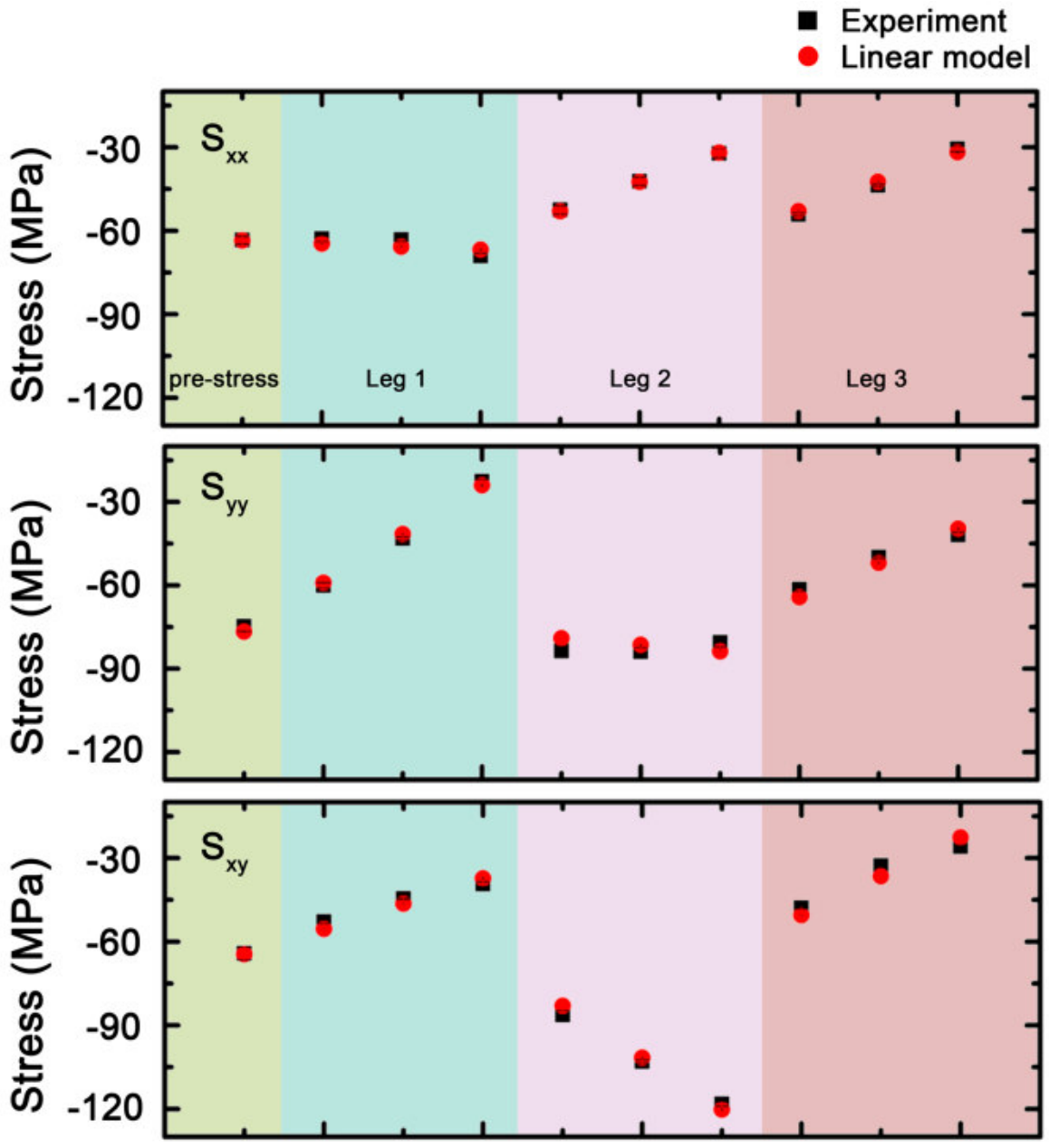

**Figure S2.-** Calculated and fitted stress tensor components. The stress tensor components obtained from the experimental PL data (black squares) are compared to the values obtained from the fitted linear transfer matrix model (red circles).

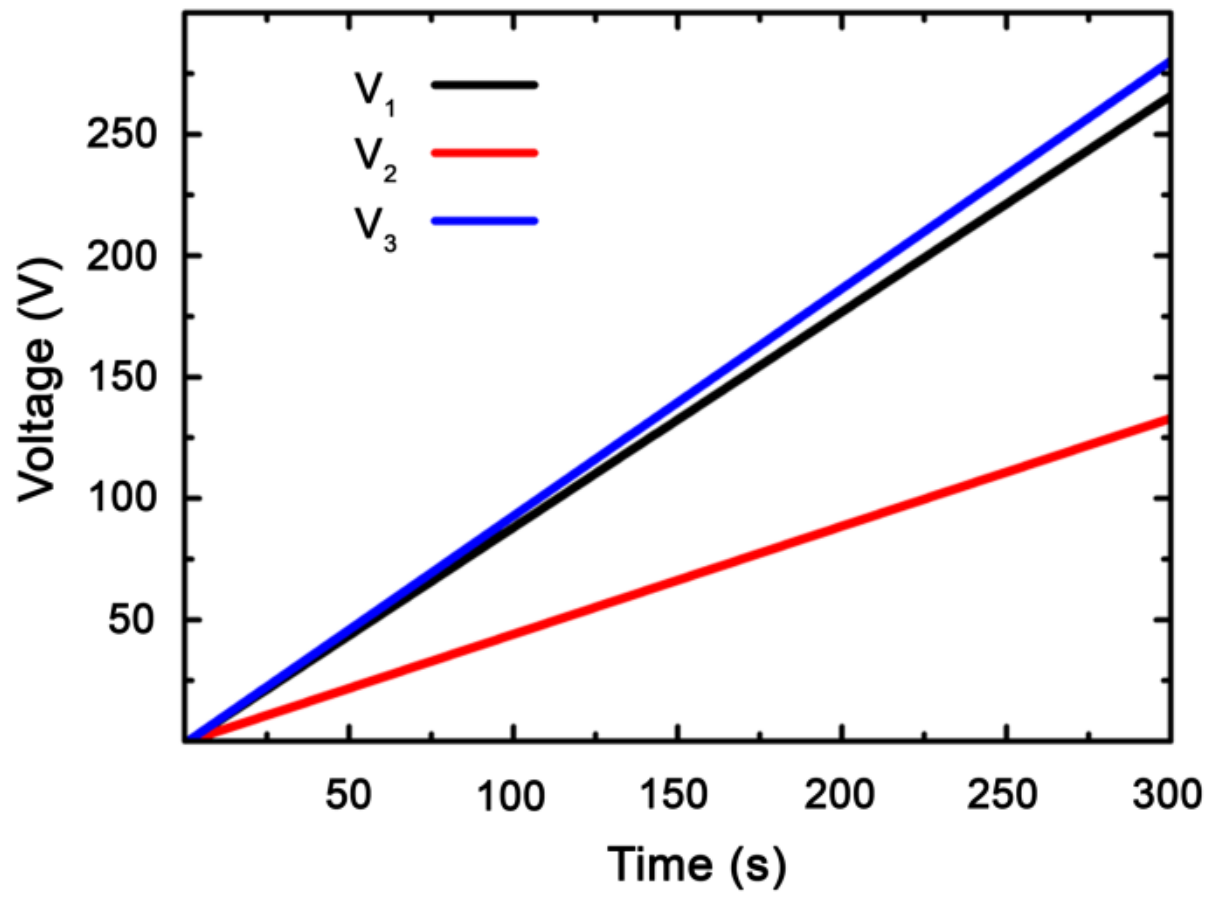

**Figure S3.-** "Programmed" voltage sweeps to exert variable uniaxial stress. The set of programmed voltages ($V_1$,$V_2$,$V_3$) to be applied to each leg are shown as a function of time in order to vary dynamically the stress from 0 to 100 MPa at a fixed stress direction $\psi_{s1}$=15°. These voltages were used to obtain the data shown in Figure 3 of the main text.

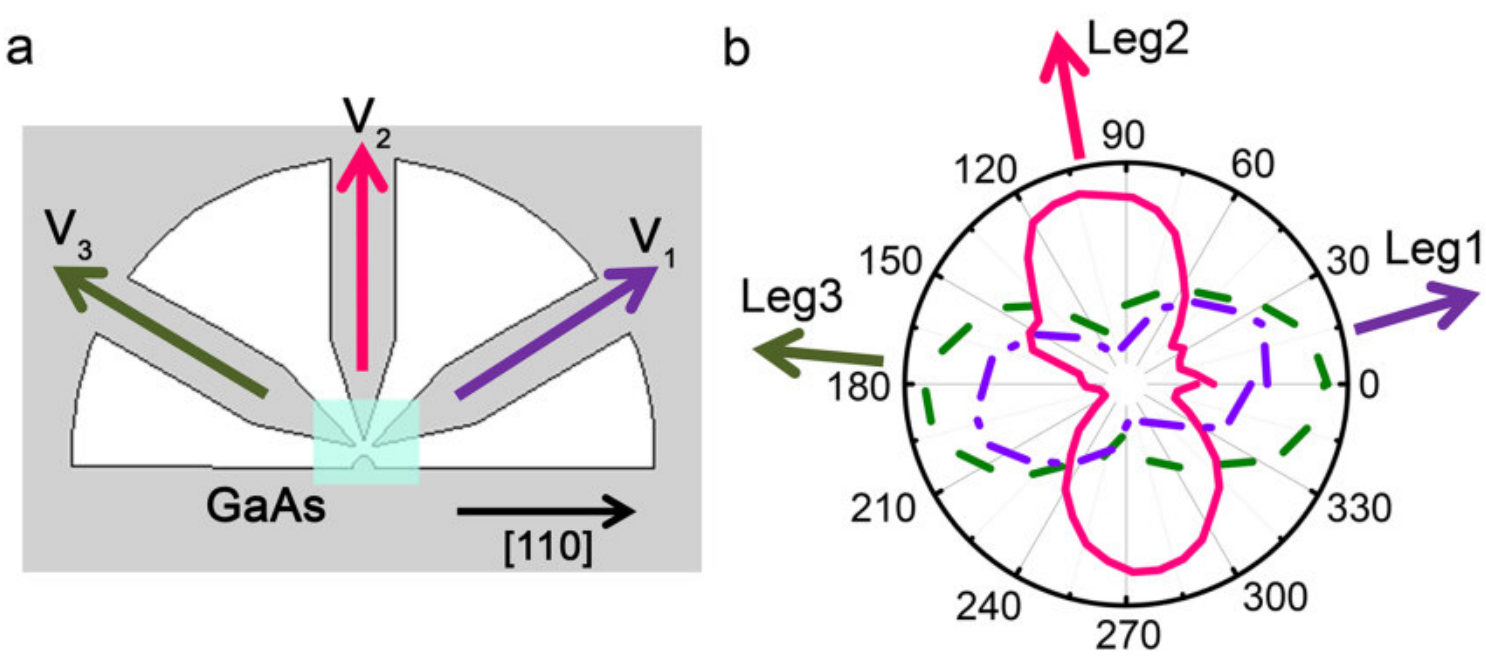

**Figure S4.-** Piezoelectric legs actuating directions. (a) Schematic representation of the piezoelectric actuator with the GaAs nanomembrane bonded on it. The PL measurements are performed close to the center of the gap between the legs. (b) The PL polarization-resolved polar plots for the high energy $E_2$ component are shown for: leg1 ($V_1$,0,0) (violet dash-dot line); leg2 (0,$V_2$,0) (pink solid line); leg3 (0,0,$V_3$) (green dash line). It is clear that the polarization direction of $E_2$ tends to align along the actuating directions extracted from the transfer matrix, confirming the validity of the model. The PL polarization-resolved measurements are referred to [110] GaAs crystallographic direction. The legs are driven individually by applying voltages $V_1$, $V_2$ and $V_3$ beginning from the unstrained state of the nanomembrane.

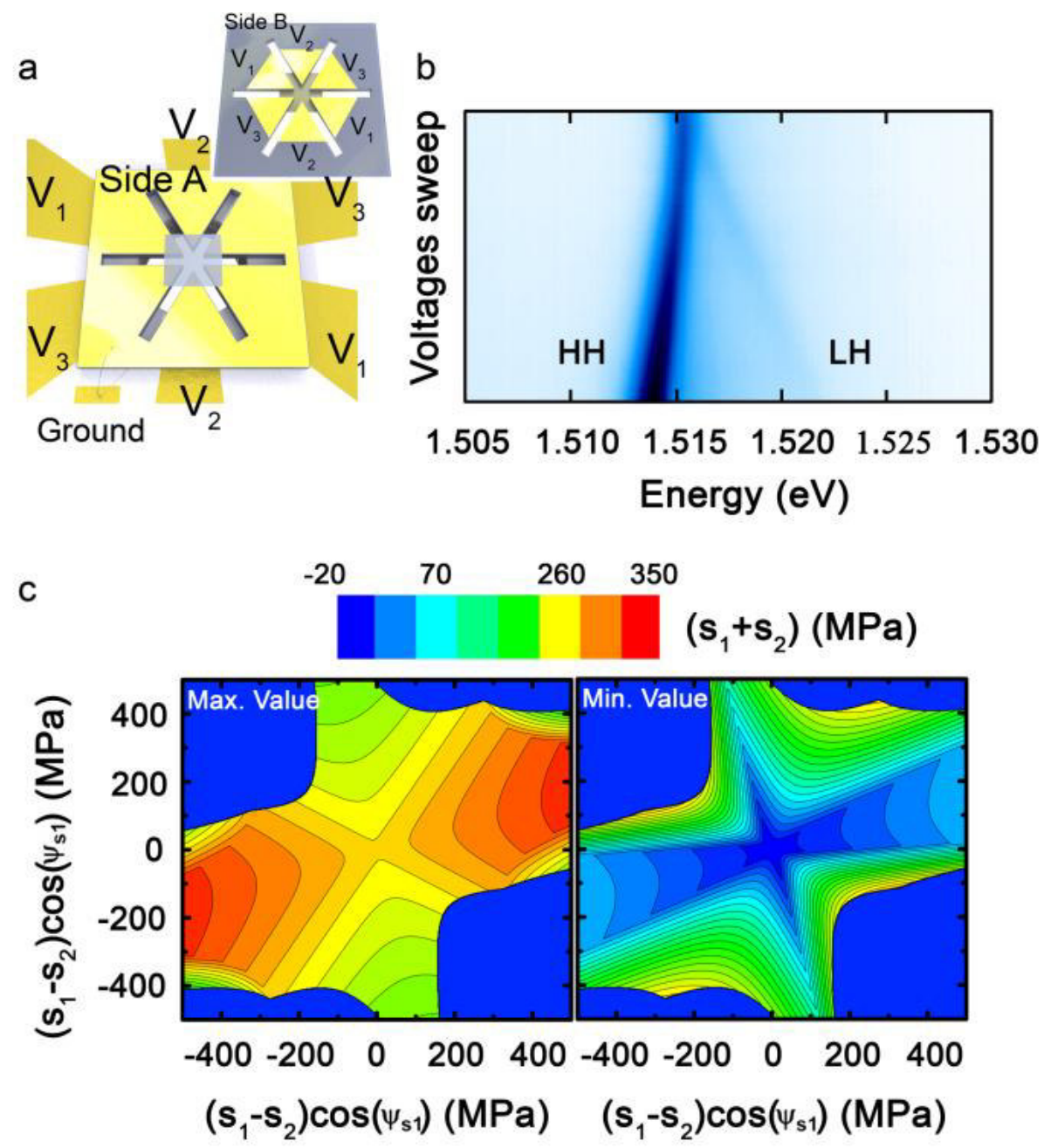

**Figure S5.-** 6-legged micro-machined actuator. (a) Sketch of the piezoelectric actuator on side A (set to ground). The GaAs nanomembrane is bonded at the center and the actuator is contacted on a AlN chip carrier through six gold pads in order to apply voltages ($V_1$, $V_2$, $V_3$) on pair of legs. The sketch of the actuator on side B where six gold contacts are patterned is shown in the inset. (b) Color-coded PL map showing LH-HH GaAs PL components shift corresponding to programmed voltages ($V_1$, $V_2$, $V_3$) in order to cancel the pre-stress state in the nanomembrane (i.e. crossing point is obtained). (c) The induced stress tuning range of the device is plotted in cylindrical coordinates for electric fields ranging from -2 to 33 kV/cm. The plots for the minimum and maximum achievable hydrostatic stressed are shown.

| **$\Gamma_1$** | **$\Gamma_2$** | **$\Gamma_3$** | **$a_c$(eV)** | **$a_g$(eV)** | **b(eV)** | **d(eV)** | **$b_c$(eV)** |
|---|---|---|---|---|---|---|---|
| 6.8 | 1.9 | 2.73 | -7.17 | -8.33 | -1.7 | -4.55 | 0 |
| **$\Delta$(eV)** | **$E_p$(eV)** | **$m_e$** | **$e_{14}$($cm^2$)** | **$a_0$(nm)** | **$C_{11}$(MPa)** | **$C_{12}$(MPa)** | **$C_{44}$(MPa)** |
| 0.341 | 25.7 | 0.067 | 0 | 0.566 | 118790 | 53760 | 59400 |

**Table S1.-** Physical parameters for GaAs used for the stress field calculation through the 8-band k.p theory. Valence and conduction band deformation potentials ($a_c$, $a_g$, b, d); spin-orbit split-off energy ($\Delta$); Kane's momentum matrix element ($E_p$); electron effective mass ($m_e$); elastic constants ($C_{11}$, $C_{12}$, $C_{44}$).